\documentclass[prd,aps,twocolumn,showpacs,nofootinbib]{revtex4-1}

\usepackage{amsfonts}
\usepackage[dvipdfm,colorlinks=true,citecolor=blue, linkcolor=blue,urlcolor=blue]{hyperref}
\usepackage{color,graphicx,amsfonts,multirow}
\usepackage{amsmath}

\begin{document}

\title{The $\rho$-meson longitudinal leading-twist distribution amplitude within QCD background field theory}

\author{Hai-Bing Fu$^{1}$}
\author{Xing-Gang Wu$^{2}$}
\email{wuxg@cqu.edu.cn}
\author{Wei Cheng$^2$}
\author{Tao Zhong$^{3}$}

\address{$^{1}$ School of Science, Guizhou Minzu University, Guiyang 550025, P.R. China}
\address{$^2$ Department of Physics, Chongqing University, Chongqing 401331, P.R. China}
\address{$^3$ Physics Department, Henan Normal University, Xinxiang 453007, P.R. China}

\date{\today}

\begin{abstract}

We revisit the $\rho$-meson longitudinal leading-twist distribution amplitude (DA) $\phi_{2;\rho}^\|$ by using the QCD sum rules approach within the background field theory. To improve the accuracy of the sum rules for its moments $\langle\xi_{n;\rho}^\|\rangle$, we include the next-to-leading order QCD correction to the perturbative part and keep all non-perturbative condensates up to dimension-six consistently within the background field theory. The first two moments read $\langle \xi_{2;\rho}^\| \rangle|_{1{\rm GeV}} = 0.241(28)$ and $\langle \xi_{4;\rho}^\| \rangle|_{1{\rm GeV}} = 0.109(10)$, indicating a double humped behavior for $\phi_{2;\rho}^\|$ at small energy scale. As an application, we apply them to the $B\to \rho $ transition form factors within the QCD light-cone sum rules, which are key components for the decay width $\Gamma(B\to \rho \ell \nu_\ell)$. To compare with the world-average of $\Gamma(B\to \rho \ell \nu_\ell)$ issued by Particle Data Group, we predict $|V_{\rm ub}| = 3.19^{+0.65}_{-0.62}$, which agrees with the BABAR and Omn\`{e}s parameterization prediction within errors.

\end{abstract}

\pacs{14.40.Be, 11.55.Hx, 12.38.Aw, 12.38.Bx}

\maketitle

\section{introduction}

The $\rho$-meson distribution amplitudes (DAs) are key components for collinear factorization of the $\rho$-meson involved processes, such as the semi-leptonic decay $B(D)\to \rho \ell \nu_{\ell}$ and the flavor-changing-neutral-current decays $B\to \rho\gamma$ and $B\to\rho \ell^+ \ell^-$, which are important for extracting the Cabibbo-Kobayashi-Maskawa (CKM) matrix elements and for searching new physics beyond the Standard Model. Inversely, those processes could provide good test of various $\rho$-meson DA models suggested in the literature. The $\rho$-meson DAs arouse people's great interest since the initial works of Refs.\cite{Lepage:1980fj, Brodsky:1981rp, Chernyak:1981zz} on the light-meson DAs. The vector $\rho$-meson DAs have more complex structures than the light pseudoscalar DAs. There are chiral-even and chiral-odd $\rho$-meson DAs due to chiral-even and chiral-odd operators in the matrix elements. The $\rho$-meson thus has two polarization states, either longitudinal ($\|$) or transverse ($\perp$), which can be expanded over different twist structures~\cite{Ball:1998sk, Ball:2004rg}. In the paper, we shall concentrate our attention on the $\rho$-meson longitudinal leading-twist DA $\phi_{2;\rho}^\|$.

The leading-twist DA $\phi_{2;\rho}^\|(x,\mu)$ can be expanded as a Gegenbauer polynomial series~\cite{Lepage:1979zb}, i.e.
\begin{equation}
\phi_{2;\rho}^{\|}(x,\mu)= 6x(1-x)\left(1+\sum_{n} C^{3/2}_{n}(\xi) \times a^\|_{n;\rho}(\mu)\right),  \label{phiDA}
\end{equation}
where $\xi=2x-1$. The Gegenbauer moments $a^\|_{n;\rho}$ at any other scale can be obtained via QCD evolution equation. The evolution equation up to next-to-leading order (NLO) is available in Ref.\cite{Ball:2006nr}. Theoretically, the Gegenbauer moments have been studied via various approaches~\cite{Ball:1996tb, Pimikov:2013usa, Ball:2007zt, Arthur:2010xf, Gao:2014bca, Forshaw:2010py, Forshaw:2012im, Choi:2007yu, Dorokhov:2006xw}. Most of their predictions are consistent with each other, but are of large theoretical uncertainties. It is helpful to provide an accurate prediction for a better comparison with the forthcoming more accurate experimental data.

The conventional Shifman-Vainshtein-Zakharov (SVZ) sum rules~\cite{Shifman:1978bx} provides a standard way to deal with the hadron phenomenology. Within the framework of SVZ sum rules, hadrons are represented by interpolating quark currents with certain quantum numbers taken at large virtualities. The correlation function (correlator) of those currents is introduced and treated by using the operator product expansion (OPE), where short- and long-distance quark-gluon interactions are separated. The short-distance part is perturbatively calculable, while the non-perturbative long-distance part can be parameterized into the non-perturbative but universal vacuum condensates. The SVZ sum rules is then achieved by matching to a sum over hadronic states with the help of dispersion relation.

The introduction of vacuum condensates is the basic assumption of the SVZ sum rules. Those universal vacuum condensates reflect non-perturbative nature of QCD which can be fixed via a global fit of experimental data. As suggested by the background-field theory (BFT), the quark and gluon fields are composed of background fields and quantum fluctuations around them. This way, the BFT provides a self-consistent description for the vacuum condensates and provides a systematic way to derive the SVZ sum rules~\cite{Novikov:1983gd, Hubschmid:1982pa, Govaerts:1983ka, Huang:1986wm, Huang:1989gv}.

The SVZ sum rules within the BFT has been applied to deal with the pseudoscalar and scalar DAs~\cite{Wang:1984rw, Xiang:1984hw, Huang:2004tp, Zhong:2011jf, Zhong:2011rg, Han:2013zg}. In those calculations, because of the complexity of high-dimensional operators and also the contribution of high-dimensional condensates are generally power suppressed, one simply adopts the reduced quark propagators $S_F(x,0)$ and the vertex operators $\Gamma (z\cdot \tensor{D})^n$, which keep only up to dimension-three operators. Such a rough treatment is theoretically incomplete, which may miss some important high-dimensional condensates in the sum rules. Their contribution may be sizable, especially to compare with the NLO QCD corrections to the perturbative part. Thus to compare with the forth-coming more and more accurate data, it is helpful to take those high-dimensional terms into consideration.

We have deduced the formulas for the quark propagator $S_F(x,0)$ and the vertex operators $\Gamma (z\cdot \tensor{D})^n$ within the BFT by keeping all terms in the OPE up to dimension-six operators~\cite{Zhong:2014jla, Zhong:2014fma}. For example, the quark propagator is parameterized as~\cite{Zhong:2014jla}
\begin{eqnarray}
S_F(x,0) &=& S_F^0(x,0)+S_F^2(x,0)+S_F^3(x,0)+ \sum_{i=1}^{2} S_F^{4(i)}(x,0) \nonumber\\
&&  + \sum_{i=1}^{3} S_F^{5(i)}(x,0) + \sum_{i=1}^{5} S_F^{6(i)}(x,0),
\end{eqnarray}
where $S_F^{k(i)}(x,0)$ stand for the propagator parts that are proportional to the dimension-$k$ operators with type $(i)$ under the same dimension. Those formulas help us to achieve a sound and accurate SVZ sum rules up to dimension-six condensates such as $\langle g_s \bar q q\rangle^2$ and $\langle g_s^3 f G^3\rangle$. Their first applications for the heavy and light pseudoscalar DAs have been done in Refs.\cite{Zhong:2014jla, Zhong:2014fma}. Those applications show that the new propagators and vertex operators shall result in new terms proportional to the dimension-six condensates that are missing in previous studies but do have sizable contributions. We shall adopt those newly derived quark propagator and vertex operators to study the $\rho$-meson longitudinal leading-twist DA $\phi_{2;\rho}^\|$. We shall then apply $\phi_{2;\rho}^\|$ to deal with the $B\to \rho$ transition form factors (TFFs) within the light-cone sum rules (LCSR)~\cite{Balitsky:1989ry, Chernyak:1990ag, Ball:1991bs}. As a further step, we shall show their effects to the $B$-meson semi-leptonic decay width $\Gamma(B\to \rho \ell \nu_\ell)$, which has been measured by the BABAR collaboration~\cite{delAmoSanchez:2010af, Aubert:2005cd}.

The remaining parts of the paper are organized as follows. In Sec.\ref{section:2}, we describe the calculation technology for deriving the moments of the leading-twist DA $\phi_{2;\rho}^\|$ within the SVZ sum rules. In Sec.\ref{section:3}, we present the numerical results for the moments $\langle \xi_{n;\rho}^\| \rangle$, the decay width $\Gamma(B\to \rho\ell\nu_\ell)$, and the CKM matrix element $|V_{\rm ub}|$. Sec.\ref{section:4} is reserved for a summary.

\section{Calculation technology}\label{section:2}

Within the framework of BFT, the gluon field $\mathcal{A}^A_\mu(x)$ and quark field $\psi(x)$ in QCD Lagrangian are replaced by
\begin{eqnarray}
\mathcal{A}^A_\mu(x) &\to& \mathcal{A}^A_\mu(x) + \phi^A_\mu(x), \label{bfrep0} \\
\psi(x) &\to& \psi(x) + \eta(x),
\end{eqnarray}
where $\mathcal{A}^A_\mu(x)$ with $A =(1, \cdots, 8)$ and $\psi(x)$ at the right-hand-side are gluon and quark background fields, respectively. $\phi^A_\mu(x)$ and $\eta(x)$ stand for the gluon and quark quantum fields, i.e., the quantum fluctuation on the background fields. The QCD Lagrangian within the BFT can be found in Ref.\cite{Huang:1989gv}. The background fields satisfy the equations of motion
\begin{equation}
(i \slash \!\!\!\! D - m)\psi(x) = 0
\end{equation}
and
\begin{equation}
\widetilde{D}^{AB}_\mu G^{B\nu\mu}(x) = g_s \bar{\psi}(x) \gamma^\nu T^A \psi(x),
\end{equation}
where $D_\mu = \partial_\mu - ig_s T^A \mathcal{A}^A_\mu(x)$ and $\widetilde{D}^{AB}_\mu = \delta^{AB} - g_s f^{ABC} \mathcal{A}^C_\mu(x)$ are fundamental and adjoint representations of the gauge covariant derivative, respectively. One can take different gauges for the quantum fluctuations and the background fields. A proper choice of gauge could make the sum rules calculation much more simplified. Practically, one usually adopts the background gauge, $\widetilde{D}^{AB}_\mu \phi^{B \mu}(x) = 0$, for the gluon quantum field~\cite{Novikov:1983gd, Hubschmid:1982pa, Govaerts:1983ka}, and the Schwinger gauge or the fixed-point gauge, $x^\mu \mathcal{A}^A_\mu(x) = 0$, for the background field~\cite{Shifman:1980ui}. Using those inputs, the quark propagator $S_F(x,0)$ and the vertex operators $\Gamma (z\cdot \tensor{D})^n$ are ready to be derived up to dimension-six operators within the BFT. We refer the interested readers to Ref.\cite{Zhong:2014jla} for details.

Considering the definition
\begin{eqnarray}
&& \langle 0|\bar d(0) z\!\!\!\slash (iz \cdot \tensor D )^n u(0) |\rho (q,\lambda ) \rangle \nonumber\\
= && (e^{(\lambda)*} \cdot z) (q \cdot z)^n m_\rho f_\rho^\parallel \langle \xi_{n;\rho}^\| \rangle,    \label{matrix:2}
\end{eqnarray}
where $q$ and $e^{(\lambda)}$ are momentum and polarization vector of the $\rho$-meson, $(z\cdot \tensor{D})^n = (z\cdot \overrightarrow{D} - z\cdot \overleftarrow{D})^n$, and $f_\rho ^\parallel$ is the decay constant. The $n_{\rm th}$-order moment of $\rho$-meson leading-twist DA $\phi_{2;\rho}^\|(x,\mu)$ at the scale $\mu$ is defined as
\begin{equation}
\langle \xi_{n;\rho}^\|\rangle = \int_0^1 du (2x-1)^n \phi_{2;\rho}^\|(x,\mu).  \label{DAmoments}
\end{equation}
As a special case, the $0_{\rm th}$-moment satisfies the normalization condition
\begin{equation}
\langle \xi_{0;\rho}^\| \rangle = \int_0^1 dx \phi_{2;\rho}^\|(x,\mu) = 1 .  \label{normalizationDA}
\end{equation}

To derive the SVZ sum rules for the $\rho$-meson leading-twist DA moments $\langle \xi_{n;\rho}^\|\rangle$, we introduce the following correlator,
\begin{eqnarray}
\Pi^{(n,0)}_\rho (z,q) &=& i \int d^4x e^{iq\cdot x}\langle 0|T \{ J_n(x) J^\dag_0(0)\}| 0 \rangle
\nonumber\\
&=& (z\cdot q)^{n+2} I^{(n,0)}(q^2),
\label{correlator}
\end{eqnarray}
where $J_n(x) = \bar{d}(x) {z\!\!\!\slash} (i z\cdot \tensor{D})^n u(x)$ and $z^2 = 0$. Here $n=(0,2,\cdots)$, i.e. only even moments are nonzero due to the isospin symmetry.

The correlator (\ref{correlator}) is an analytic $q^2$-function defined at both positive and negative $q^2$-values. In physical region ($q^2>0$), the complicated hadronic content of the correlator can be quantified by applying the unitarity relation through inserting a complete set of intermediate hadronic states to the matrix element. By singling out the ground-state and introducing a compact notation for the rest of contributions, we obtain
\begin{eqnarray}
\textrm{Im}  I^{(n,0)}_{\rm had}(q^2) & =& \pi \delta (q^2 - m_\rho^2) {f_\rho^\|}^2 \langle \xi_{n;\rho}^\|\rangle  \nonumber\\
&&+ \pi \frac{3}{4\pi^2 (n+1) (n+3)} \theta (q^2 - s_\rho), \label{hadim}
\end{eqnarray}
where the quark-hadron duality has been adopted and the parameter $s_\rho$ is the continuum threshold of the lowest continuum state.

\begin{figure}[htb]
\includegraphics[width=0.40\textwidth]{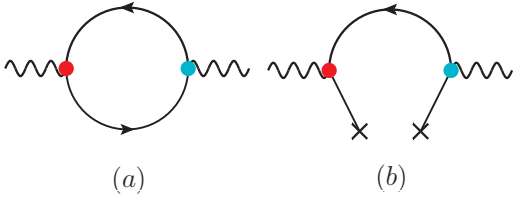}
\caption{Schematic Feynman diagrams for the $\rho$-meson longitudinal leading-twist DA moments, where the cross $(\times)$ stands for the background quark field. The big dots stand for the vertex operators in the correlator, the left one is for $z\!\!\!\slash (iz \cdot \tensor D )^n$ and the right one is for $z\!\!\!\slash$.}
\label{feyn}
\end{figure}

On the other hand, one can apply the OPE for the correlator (\ref{correlator}) in deep Euclidean region $(q^2<0)$. The coefficients before the operators (result in non-perturbative condensates) are perturbatively calculable. The OPE indicates that
\begin{eqnarray} &&
\Pi^{(n,0)}_{2;\rho} (z,q) = i \int d^4x e^{iq\cdot x}
\nonumber\\
&& \quad \times \big\{ - {\rm Tr} \langle 0 |  S^d_F(0,x) z\!\!\!\slash  (iz\cdot \tensor{D})^n S^u_F(x,0) z\!\!\!\slash | 0\rangle
\nonumber\\
&& \quad  + {\rm Tr} \langle 0 |  \bar{d}(x) d(0) z\!\!\!\slash  (iz\cdot \tensor{D})^n S^u_F(x,0) z\!\!\!\slash | 0\rangle
\nonumber\\
&& \quad  + {\rm Tr} \langle 0 |  S^d_F(0,x) z\!\!\!\slash (iz\cdot \tensor{D})^n \bar{u}(0) u(x) z\!\!\!\slash | 0\rangle  \big\} + \cdots ~.
\label{ope}
\end{eqnarray}
The first term corresponds to Fig.(\ref{feyn}a), the second one corresponds to Fig.(\ref{feyn}b), and the third one corresponds to permutation contribution by transforming $u\leftrightarrow d$ from Fig.(\ref{feyn}b). We adopt the dimensional regularization under the $\overline{\rm MS}$-scheme to deal with the infrared divergences at high orders, whose divergent terms shall be absorbed into the renormalized leading-twist DA~\cite{Li:2012gr}.

As a combination of the correlator within different $q^2$-region, the sum rules for the moments of the $\rho$-meson leading-twist DA can be derived by using the dispersion relation
\begin{eqnarray}
\frac{1}{\pi} \frac{1}{M^2} \int ds e^{-s/M^2} \textrm{Im} I_{\rm had}(s) = \hat{L}_M  I_{\rm QCD}(q^2), \label{bordisrel}
\end{eqnarray}
where $M$ is the Borel parameter and the Borel transformation operator
\begin{eqnarray}
\hat{L}_M = \lim_{\begin{array}{c} Q^2,n\to\infty\\ Q^2/n=M^2 \end{array}} \frac{1}{(n-1)!} (Q^2)^n \left( -\frac{d}{dQ^2} \right)^n,
\end{eqnarray}
where $Q^2=-q^2$. The final sum rules reads
\begin{widetext}
\begin{eqnarray}\label{xiSR}
&&\langle \xi_{n;\rho}^\| \rangle  = \frac{M^2}{f_\rho^2}e^{m_\rho^2/M^2} \Bigg\{ \frac{3}{4\pi^2(n+1)(n+3)}\left(1 + \frac{{{\alpha _s}}}{\pi }A_n'\right) \left( 1 - e^{-s_\rho/M^2} \right) + \sum\limits_{q=u,d} \left( \frac{m_q \langle \bar q q\rangle}{M^4}  - \frac{8n+1}{18} \frac{m_q \langle g_s \bar q\sigma TGq\rangle}{M^6}\right.
\nonumber
\\
&&\quad + \left.\frac{4n+2}{81} \frac{\langle g_s\bar q q \rangle^2}{M^6}\right)+ \frac{1 + n\theta(n-2)}{12\pi(n+1)} \frac{\langle \alpha_s G^2\rangle }{M^4}  + \frac{1}{16\pi}\frac{\langle g_s^3fG^3\rangle}{M^6} \left\{\frac{8\delta^{n0} + 405n + 192}{36}\ln\frac{M^2}{\mu^2} - \frac{16\delta^{n0} + 810n + 363}{72}  \right.
\nonumber
\\
&& \quad \times \gamma_E+ \frac{7}{24} \psi(n+1) + \frac{8 \delta^{n0} + 405n + 826}{72} + \theta (n - 2)\left[ \frac{16 - 22n}{72}\ln \frac{M^2}{\mu^2} - \frac{788n + 421}{72}\psi(n + 1) - \frac{766n + 437}{72} \gamma_E  \right.
\nonumber
\\
&&\quad - \frac{68n^2 - 37n - 11}{144n} + \left.\left. \sum\limits_{k=0}^{n-2} {(-1)^k} \frac{1}{144}\left( \frac{3(135k + 128)}{n-k} + \frac{383k+399}{k-n+1} - \frac{106kn-410k+617n-415}{(k + 1)(k + 2)} + 106 \right) \right]\right\}\Bigg\}, \nonumber\\
\label{momsr}
\end{eqnarray}
\end{widetext}
where the step function $\theta(x)=1$ for $x\geq0$, and $\theta(x)=0$ for $x<0$. $\gamma_E=0.557216$ is Euler's constant and the $0_{\rm th}$-derivative of the digamma function $\psi(n+1)=\sum_{k=1}^n 1/k -\gamma_E$. The NLO coefficients $A_n$ have been calculated by Ref.\cite{Ball:1996tb}, whose first three ones are, $A'_0 = 1$, $A'_2 = \frac{5}{3}$ and $A'_4 = \frac{59}{27}$, respectively.

One can obtain relations among the Gegenbauer moments $a_{n;\rho}^\|$ and the moments $\langle \xi_{n;\rho}^\|\rangle$ by substituting Eq.(\ref{phiDA}) into Eq.(\ref{DAmoments}). For examples, we have
\begin{eqnarray}
a_{2;\rho}^\| &=& \frac{7}{12} \left( 5\langle \xi^\|_{2;\rho} \rangle  - 1 \right),  \label{gegmom1} \\
a_{4;\rho}^\| &=& -\frac{11}{24} \left( 14\langle \xi^\|_{2;\rho} \rangle  - 21\langle \xi_{4;\rho}^\| \rangle  - 1 \right), \label{gegmom2} \\
a_{6;\rho}^\| &=& \frac{5}{64} \left( 135\langle \xi^\|_{2;\rho} \rangle - 495\langle \xi^\|_{4;\rho} \rangle + 429\langle \xi^\|_{6;\rho} \rangle - 5 \right). \label{gegmom3}
\end{eqnarray}

\section{Numerical results and discussions} \label{section:3}

We adopt the following parameters to do the numerical calculation. The $\rho$-meson mass and decay constant are from the Particle Data Group~\cite{Agashe:2014kda}, $m_\rho = 0.775 {\rm GeV}$ and $f_\rho^\| = 0.216 \pm 0.003 {\rm GeV}$. The non-perturbative vacuum condensates up to dimension-six have been determined in references~\cite{Nambu:1960xd, GellMann:1968rz, Narison:2010wb, Belyaev:1982sa, Narison:2009vy, Bordes:1988yr, Dominguez:1987nw, Causse:1989cr, Launer:1983ib, Bertlmann:1984ih, Narison:1995jr, Narison:2001ix, Chung:1981wm},
\begin{eqnarray}
\langle\bar q q\rangle &=& -0.0138(17){\rm GeV}^3, \nonumber\\
\langle g_s \bar q q\rangle^2 &=& -0.0018(7){\rm GeV}^6, \nonumber\\
\langle\alpha_s G^2\rangle &=& 0.038(11){\rm GeV}^4, \nonumber\\
\langle g_s^3 f G^3\rangle &=& 0.013(7){\rm GeV}^6, \nonumber\\
\sum_{q=u,d} m_q \langle \bar{q}q\rangle &=& -1.656(5)\times 10^{-4} {\rm GeV}^4, \nonumber\\
~\sum_{q=u,d}m_q \langle g_s\bar{q}\sigma TGq\rangle &=& 1.325(33) \times 10^{-4} \textrm{GeV}^4. \nonumber
\end{eqnarray}

The continuum threshold $s_\rho$ is usually set as the value around the squared mass of the $\rho$-meson first excited state. At present, the structure of the excited $\rho$-meson state is not yet completely clear, cf. a recent review in Ref.\cite{Agashe:2014kda}. Therefore, we use the sum rules (\ref{momsr}) with $n=0$, together with the normalization condition $\langle \xi_{0;\rho}^\|\rangle=1$, to inversely determine an effective value for $s_\rho$. We get, $s_\rho\simeq 2.8{\rm GeV}^2$, which indicates that the effective threshold continuum state is close to $\rho(1700)$.

\subsection{The $\rho$-meson leading-twist DA $\phi_{2;\rho}^\|(x,\mu)$}

To determine a Borel window for the sum rules (\ref{momsr}), e.g. the allowable range for $M$, we adopt two criteria: I) All continuum contributions are less than $40\%$ of the total dispersion relation; II) The contributions from the dimension-six condensates should not exceed $10\%$. By setting all other parameters to be their central values, the first two moments $\langle\xi_{2;\rho}^\|\rangle$ and $\langle\xi_{4;\rho}^\|\rangle$ up to NLO level at the scale $\mu=M$ are determined as
\begin{equation}
\langle\xi_{2;\rho}^\|\rangle|_{\mu=M}=0.234(23) \;\;{\rm for}\;\; M^2\in [1.72,3.00]
\end{equation}
and
\begin{equation}
\langle\xi_{4;\rho}^\|\rangle|_{\mu=M}=0.103(7) \;\;{\rm for}\;\; M^2\in [4.26,4.86],
\end{equation}
where the central values are for $M^2=2.185$ and $4.535$, respectively.

\begin{widetext}
\begin{center}
\begin{table}[htb]
\caption{The first two moments $\langle\xi_{(2,4);\rho}^\|\rangle|_{\mu=M}$ of the longitudinal leading-twist DA $\phi_{2;\rho}^\|$ predicted from the sum rules under the BFT. Here the perturbative contributions are calculated up to NLO level and the non-perturbative contributions are up to dimension-six condensates. The contributions from the LO-terms, the NLO-terms, the dimension-three, the dimension-four, the dimension-five and the dimension-six condensates are presented separately. The errors are obtain by varying $M^2$ within the determined Borel window. }
\begin{tabular}{c| c c c c c cc }
\hline                                                  & LO  & NLO & Dimension-three & Dimension-four & Dimension-five & Dimension-six & Total \\ \hline
$\langle \xi_{2;\rho}^\| \rangle |_{\mu=M} $ & 0.193(34)& $0.014(1)$ & $-0.0021(7)$ & 0.013(6) & 0.0007(5) & 0.015(22) & 0.234(23) \\
$\langle \xi_{4;\rho}^\| \rangle |_{\mu=M} $ & 0.081(4) & $0.008(1)$ & $-0.0008(1)$ & 0.005(2) & 0.0003(1)& 0.010(1) & 0.103(7) \\ \hline
\end{tabular}\label{NLO6dim}
\end{table}
\end{center}
\end{widetext}

To show how the non-perturbative dimension-six condensates and the perturbative NLO corrections affect the moments, we list the first two moments $\langle\xi_{(2,4);\rho}^\|\rangle$ at the scale $M$ in Table \ref{NLO6dim}, where the perturbative contributions are calculated up to NLO level and the non-perturbative contributions are up to dimension-six condensates. The contributions from the LO-terms, the NLO-terms, the dimension-three, the dimension-four, the dimension-five and the dimension-six condensates are presented separately in Table \ref{NLO6dim}. It shows that the dominant contribution is from the LO-terms, which provide $\sim80\%$ contribution to $\langle\xi_{(2,4);\rho}^\|\rangle$. The NLO-terms provide $\sim 6.0\%$ contribution to $\langle\xi_{2;\rho}^\|\rangle$ and $\sim 7.8\%$ contribution to $\langle\xi_{4;\rho}^\|\rangle$. It is noted that the non-perturbative condensates do not follow the usual power counting of $1/M^2$-suppression, and the dimension-six condensates provide sizable contributions to the moments which are at the same order of the NLO-terms. Thus they are of equal importance for a precise prediction of the $\phi_{2;\rho}^\|$ moments.

By using the relations among the Gegenbauer moments $a_{n;\rho}^\|$ and the moments $\langle \xi_{n;\rho}^\|\rangle$, such as Eqs.(\ref{gegmom1},\ref{gegmom2},\ref{gegmom3}), we can derive $a_{n;\rho}^\|$ at the scale $M$. Furthermore, the Gegenbauer moments $a_{n;\rho}^\|$ at any other scale can be obtained via the QCD evolution, i.e. the evolution at the NLO accuracy shows~\cite{Floratos:1977au, Mueller:1994cn, Ball:2006nr}
\begin{eqnarray}
a^{\|}_{n;\rho}(\mu) &=&  a_{n;\rho}^\|(\mu_0) E_{n;\rho}^{\rm NLO} \nonumber\\
&+& \frac{\alpha_s(\mu)}{4\pi}\sum_{k=0}^{n-2} a_{k;\rho}(\mu_0)\,
L^{\gamma_k^{(0)}/(2\beta_0)}d^{(1)}_{nk},
\end{eqnarray}
where $\mu_0$ is the initial scale, $\mu$ is the required scale, and
\begin{eqnarray}
E_{n;\rho}^{\rm NLO} &=&  L^{\gamma^{(0)}_n/(2\beta_0)} \nonumber\\
&& \times\left\{1+ \frac{\gamma^{(1)}_n \beta_0-\gamma_n^{(0)}\beta_1}{8\pi\beta_0^2}
\Big[\alpha_s(\mu)-\alpha_s(\mu_0)\Big]\right\},
\end{eqnarray}
where $L=\alpha_s(\mu)/\alpha_s(\mu_0)$, $\beta_0=11-2n_f/3$, $\beta_1=102-38n_f/3$, $\gamma_n^{(0)}$ and $\gamma_n^{(1)}$ are LO and NLO anomalous dimensions, accordingly.

\begin{widetext}
\begin{center}
\begin{table}[htb]
\caption{The first two Gegenbauer moments $a_{2;\rho}^\|$ and $a_{4;\rho}^\|$ for the longitudinal leading-twist DA $\phi_{2;\rho}^\|$, which is predicted from the sum rules under the BFT. A comparison of predictions under various approaches~\cite{Pimikov:2013usa, Ball:2007zt, Arthur:2010xf, Gao:2014bca, Forshaw:2010py, Forshaw:2012im, Choi:2007yu, Dorokhov:2006xw} has also been presented. For easy comparison, we have set the scale $\mu_{0}=1{\rm GeV}$. The moments $\langle\xi_{2;\rho}^\|\rangle$ and $\langle\xi_{4;\rho}^\|\rangle$ and the inverse moment $\langle x^{-1}\rangle$ are also presented. The number in the parenthesis shows the uncertainties from all the input parameters. }
\begin{tabular}{c| c c c c c}
\hline
~ ~~~~~~&~~~~~~ $a_{2;\rho}^\|$ ~~~~~~&~~~~~~ $a_{4;\rho}^\|$ ~~~~~~&~~~~~~ $\langle \xi_{2;\rho}^\| \rangle$ ~~~~~~&~~~~~~ $\langle \xi_{4;\rho}^\| \rangle$ ~~~~~~&  $\langle x^{-1} \rangle$ \\
\hline
our predictions &    $0.119(82)$   & $-0.035(100)$       &  $0.241(28)$       &  $0.109(10)$      &  3.30(34)   \\
NLCSR~\cite{Pimikov:2013usa}                  &    0.047(58)     &  $-0.057(118)$      &  0.216(21)         & 0.089(9)         &  2.97(39)   \\
BB~\cite{Ball:2007zt}                         &    0.150(70)     & -                   &  0.251(24)         & -        &  3.45(21)   \\
Lattice QCD~\cite{Arthur:2010xf}              &    0.197(158)    & -                   &  0.268(54)         & -        &  3.60(48)   \\
BS~\cite{Gao:2014bca}                &    0.111         &    0.036            &  0.238             & 0.115            &  3.44       \\
AdS/QCD~\cite{Forshaw:2010py, Forshaw:2012im}  &    0.104         & 0.053               & 0.236              & 0.115            & 3.47        \\
LFQM~\cite{Choi:2007yu}                       &    0.014         & $-0.005$            & 0.205              & 0.088            & 3.03       \\
IM~\cite{Dorokhov:2006xw}                     &    $-0.010$      & $-0.033$            & $0.196$            & $0.080$          & 2.87        \\
\hline
\end{tabular}\label{taba2}
\end{table}
\end{center}
\end{widetext}

We present our predictions for the Gegenbauer moments $a_{2(4);\rho}^\|$, together with the moments $\xi_{2(4);\rho}^\|$ and the inverse moment $\langle x^{-1} \rangle = \int_0^1 dx x^{-1} \phi_{2;\rho}^\| (x,\mu)$, in Table~\ref{taba2}, where all uncertainty sources have been taken into consideration and have been summed up in quadrature. Because of the dominance of the LO-terms to the moments $\langle\xi_{(2,4);\rho}^\|\rangle$, the $\phi_{2;\rho}^\|$ behavior and the quantities such as the TFFs and $|V_{\rm ub}|$ shall be dominated by the LO-terms. For example, as will be shown later, if without taking the NLO-terms and the dimension-six condensates into consideration, the magnitudes of $A_1$, $A_2$ and $V_{0}$ at $q^2=0$ shall be altered by $3\%-4\%$; and the magnitude of $|V_{ub}|$ shall be altered by $\sim5\%$.

\begin{figure}[htb]
\centering
\includegraphics[width=0.45\textwidth]{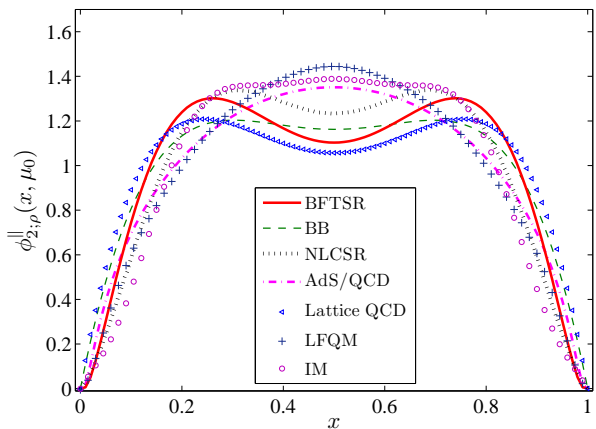}
\caption{The $\rho$-meson leading-twist DA $\phi_{2;\rho}^\|(x,\mu_{0} =1{\rm GeV})$ predicted from the sum rules under the BFT (BFTSR). As a comparison, the NLCSR prediction~\cite{Pimikov:2013usa}, the BB prediction~\cite{Ball:2007zt}, the Lattice QCD prediction~\cite{Arthur:2010xf}, the AdS/QCD prediction~\cite{Forshaw:2010py, Forshaw:2012im}, the LFQM prediction~\cite{Choi:2007yu}, and the IM prediction~\cite{Dorokhov:2006xw} have also been presented. } \label{DA:comparation}
\end{figure}

As a comparison, we also present the sum rules prediction with nonlocal condensates (NLCSR)~\cite{Pimikov:2013usa}, the Ball and Brawn (BB) prediction~\cite{Ball:2007zt}, the Lattice QCD prediction~\cite{Arthur:2010xf}, the Bethe-Salpeter wavefunction (BS) prediction~\cite{Gao:2014bca}, the AdS/QCD prediction~\cite{Forshaw:2010py, Forshaw:2012im}, the Light-Front Quark Model (LFQM) prediction~\cite{Choi:2007yu}, and the Instanton Model (IM) prediction~\cite{Dorokhov:2006xw} in Table~\ref{taba2}. To compare with the other predictions, we have set the scale $\mu_{0}=1{\rm GeV}$, which is adopted in most of the references. It is noted that our present predictions on the DA moments agree with most of them within reasonable errors, and most of them prefer a double humped behavior, as explicitly shown by Fig.(\ref{DA:comparation}).

\begin{figure}[htb]
\includegraphics[width=0.45\textwidth]{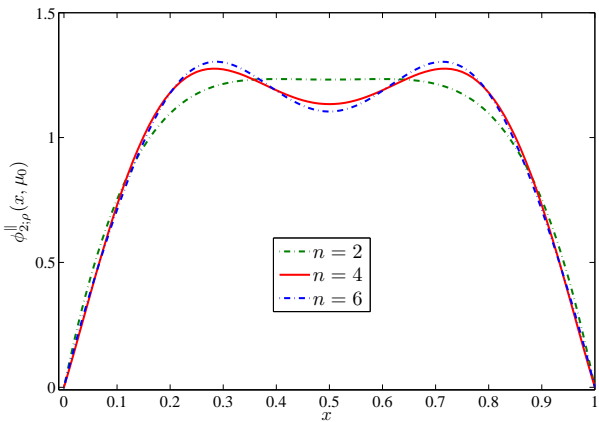}
\caption{The $\rho$-meson longitudinal twist-2 DA $\phi_{2;\rho}^\|(x,\mu_0=1{\rm GeV})$ for $n =2$, $4$ and $6$, respectively.}
\label{fig2}
\end{figure}

We make a discussion on how the $\phi_{2;\rho}^\|$ behavior changes with different truncations of the Gegenbauer expansion. By taking the central values for the Gegenbauer moment $a_{n;\rho}^\|$, we put the DA $\phi_{2;\rho}^\|(x,\mu_0=1{\rm GeV})$ for $n=(2,4,6)$ in Fig.(\ref{fig2}). It shows that by including the sixth-moment $a_{6;\rho}^\| (1{\rm GeV})= 0.009 $ into the Gegenbauer expansion, the shape of $\phi_{2;\rho}^\|$ is slightly changed and close to the double humped behavior for the case of $n=4$. By including more moments into the expansion, the $\phi_{2;\rho}^\|$ behavior shall be almost unchanged. Thus it is convenient and is of high precision to keep only the first two moments in the Gegenbauer expansion.

The $\rho$-meson leading-twist wavefunction $\psi_{2;\rho }^\| (x,\mathbf{k}_\perp)$ is an important component for a reliable pQCD predictions within the $k_T$ factorization formalism~\cite{Botts:1989kf, Li:1992nu}. We adopt the present DA moments to fix a $\rho$-meson wavefunction $\psi_{2;\rho}^\|(x,{\bf k}_\bot)$ that is constructed from the Wu-Huang (WH) model~\cite{Wu:2010zc}
\begin{eqnarray}
\psi_{2;\rho}^\|(x,{{\bf{k}}_ \bot }) = \sum\limits_{{h _1}{h _2}} {{\chi_\rho^{{h _1}{h_2}}}} (x,{{\bf{k}}_ \bot }) \psi _{2;\rho}^{R}(x,{{\bf{k}}_\bot }),
\end{eqnarray}
whose radial part is from the BHL-prescription~\cite{BHL}. The spin-space wavefunction ${{\chi_\rho^{{h _1}{h_2}}}} (x,{{\bf{k}}_ \bot })$ is from the Wigner-Melosh rotation~\cite{Huang:1994dy, Cao:1997hw, Huang:2004su}. The $\rho$-meson DA $\phi^\|_{2;\rho}$ can be derived from $\psi_{2;\rho}^\|(x,{{\bf{k}}_ \bot })$ via the relation
\begin{eqnarray}
\phi_{2;\rho}^\|(x,\mu) = \frac{ 2\sqrt{3}}{ \widetilde{f}_\rho^\|}\int_{|{\bf k}_\bot|^2\leq\mu^2}\frac{d{\bf k}_\bot}{16\pi^3}\psi_{2;\rho}^\|(x,{\bf k}_\bot)\,,
\end{eqnarray}
which leads to
\begin{eqnarray}
&&\phi _{2;\rho }^\| (x,\mu)  = \frac{{A_{2;\rho}^\| \sqrt {3x\bar x} {m_q}}}{{8{\pi ^{\frac{3}{2}}}\widetilde f_\rho ^\| b_{2;\rho}^\| }}[1 + {B_{2;\rho}^\| }C_2^{\frac{3}{2}}(\xi )+ C_{2;\rho}^\| C_4^{\frac{3}{2}}(\xi )]\nonumber\\
&&\qquad\times  \left[ {{\rm{Erf}}\left( {b_{2;\rho}^\| \sqrt {\frac{{{\mu^2} + m_q^2}}{{x\bar x}}} } \right) - {\rm{Erf}}\left( {b_{2;\rho}^\| \sqrt {\frac{{m_q^2}}{{x\bar x}}} } \right)} \right], \nonumber\\ \label{DA:WH}
\end{eqnarray}
where $\widetilde{f}_\rho^\| = f_\rho^{\|}/\sqrt{5}$, the error function, $\textrm{Erf}(x) = 2 \int^x_0 e^{-t^2} dt/ \sqrt \pi$ and the light constitute quark mass, $m_q\simeq300$ GeV. To be slightly different from the one suggested in Ref.\cite{Fu:2014cna}, we have explicitly put the newly derived fourth Gegenbauer term in the longitudinal function. Four model parameters can be fixed by the normalization condition, the average value of the squared transverse momentum $\langle {\bf k}_\bot^2 \rangle_{2;\rho}^{1/2} = 0.37 \pm 0.02 {\rm GeV}$, and the second and fourth Gegenbauer moments determined from the sum rules (\ref{momsr}). By using the central values for the input parameters, we obtain: $A_{2;\rho}^\|=24.61$, $b_{2;\rho}^\|=0.581$, $B_{2;\rho}^\| = 0.075$ and $C_{2;\rho}^\| = -0.044$.

\subsection{The $B\to \rho$ transition form factors}

\begin{figure}[htb]
\centering
\includegraphics[width=0.45\textwidth]{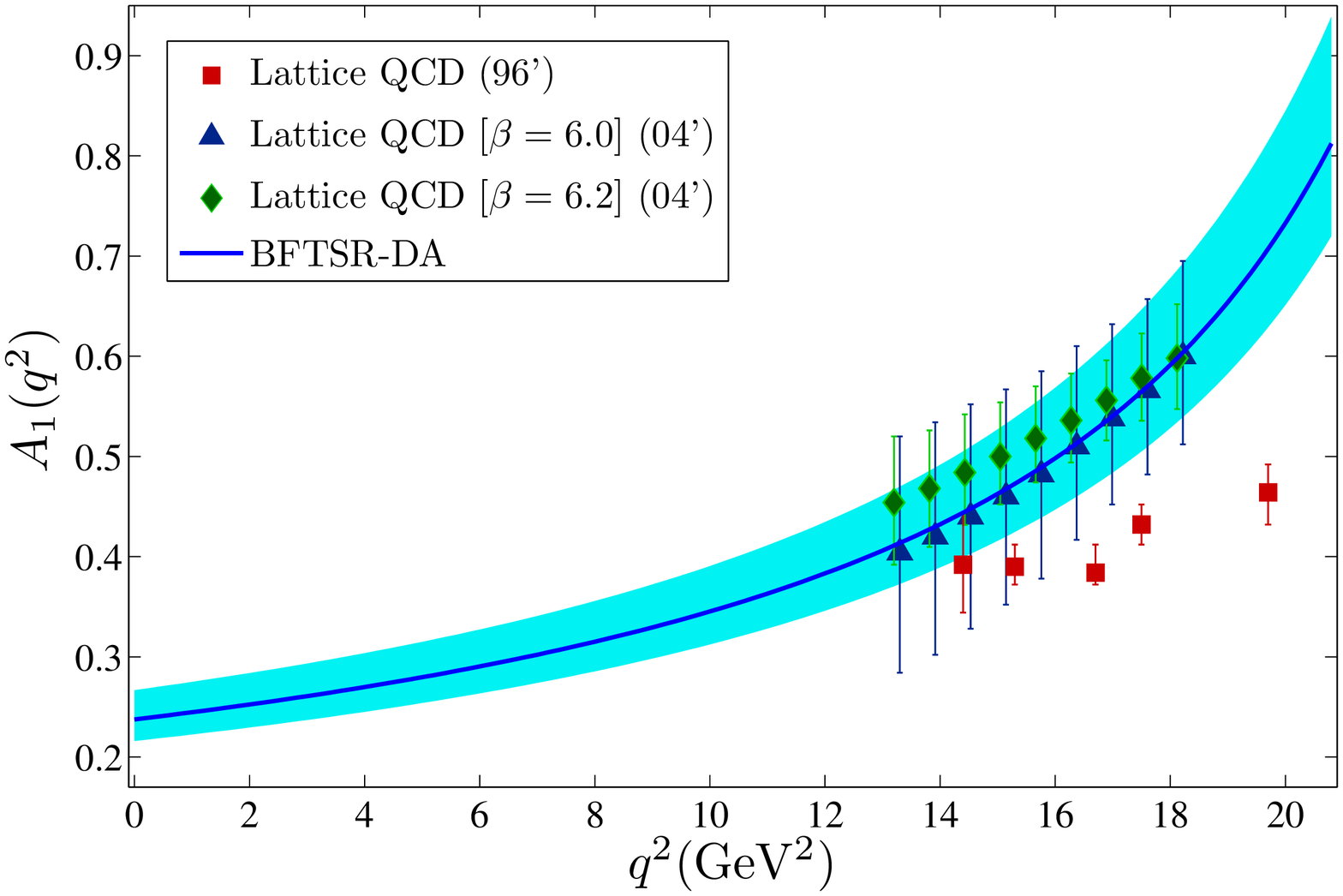}
\includegraphics[width=0.45\textwidth]{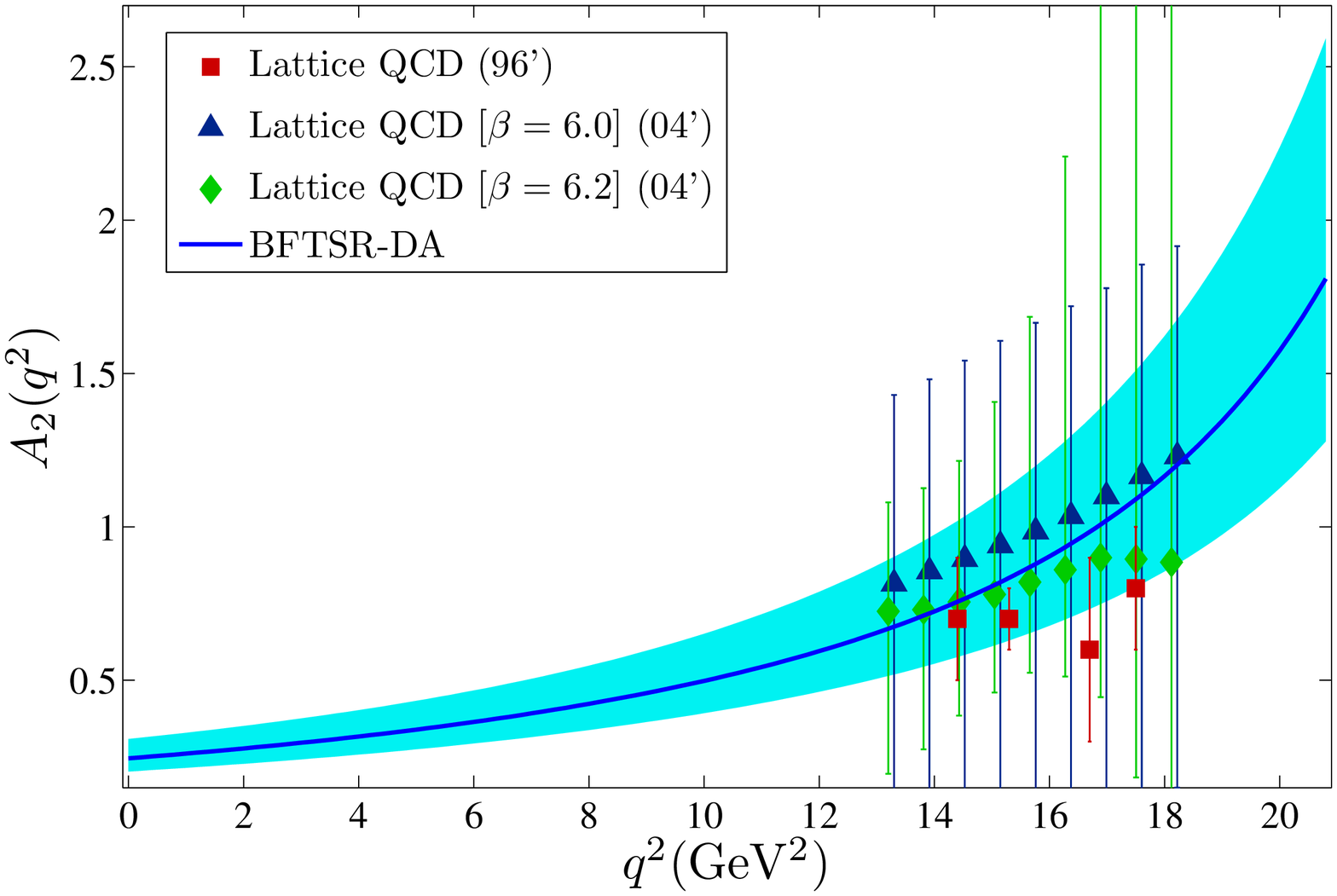}
\includegraphics[width=0.45\textwidth]{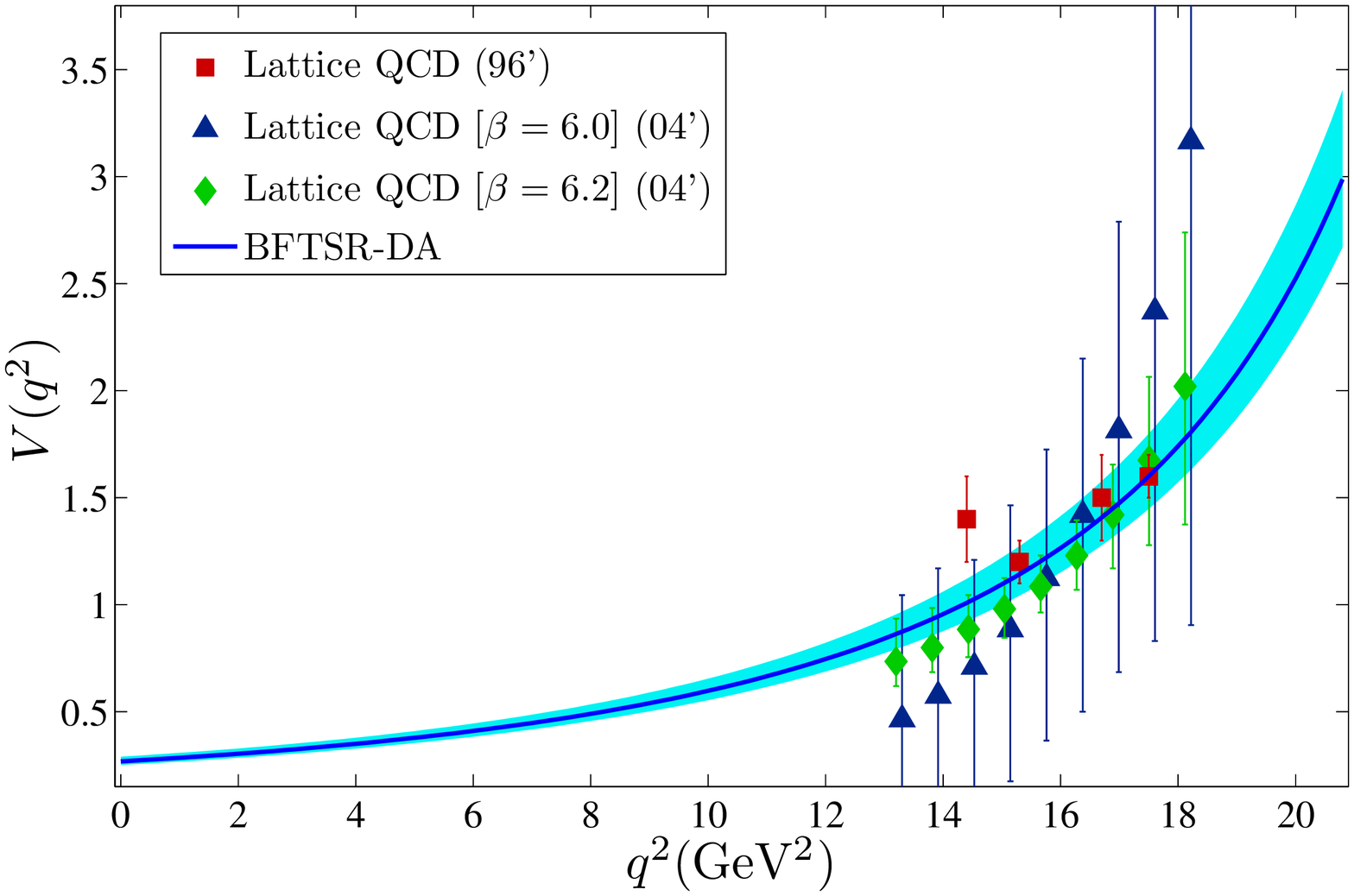}
\caption{The extrapolated $B\to \rho$ axial-vector and vector TFFs $A_{1,2}(q^2)$ and $V(q^2)$ by using the LCSRs derived in Ref.\cite{Fu:2014cna}, where the $\rho$-meson leading-twist DA $\phi_{2;\rho}^\|$ is from our present sum rules under the BFT (BFTSR). The lattice QCD predictions~\cite{Flynn:1995dc, Bowler:2004zb} are presented as a comparison. } \label{fig_TFF}
\end{figure}

\begin{table}[htb]
\begin{tabular}{ c  c c c }
\hline
          & ~~~~$A_1$~~~~ & ~~~~$A_2$~~~~ & ~~~~$V$~~~~
\\ \hline
$a_1^i$   & 0.233 & $-0.874$ & $-1.034$ \\
$a_2^i$   & 0.345 & $0.708$ & $5.257$ \\
$\Delta$  &  0.16 &   0.23   & 0.41     \\
\hline
\end{tabular}
\caption{The fitted parameters $a^i_{1,2}$ for the $B\to \rho$ TFFs $F_i$, in which all the LCSR parameters are set to be their central values. $\Delta$ is the measure of the quality of extrapolation. } \label{analytic}
\end{table}

One of the important application of $\phi_{2;\rho}^\|$ is the $B$-meson semi-leptonic decay $B\to\rho\ell\nu_\ell$. It is the key component for the vector and axial vector $B\to\rho$ TFFs $A_{1}(q^2)$, $A_{2}(q^2)$ and $V(q^2)$. By using a left-handed current $j_B^\dag (x) = i\bar b(x)(1-\gamma_5)q_2(x)$ to do the LCSR calculation, one can highlight the contributions from $\phi_{2;\rho}^\|$~\cite{Fu:2014cna}, thus showing the properties of $\phi_{2;\rho}^\|$ via a more transparent way. Following the standard LCSR procedures, one can derive the LCSRs for the mentioned TFFs, which have been presented in Ref.\cite{Fu:2014cna}. One only needs to replace the DA $\phi_{2;\rho}^\|$ used there to be our present one.

At the large recoil region, $q^2 \approx 0 {\rm GeV}^2$, we obtain
\begin{eqnarray}
A_1(0) &=& 0.237^{+0.029}_{-0.021}~, \label{F01} \\
A_2(0) &=& 0.246^{+0.063}_{-0.043}~, \label{F02} \\
V(0)   &=& 0.268^{+0.021}_{-0.017}~, \label{F03}
\end{eqnarray}
where the errors are squared averages of all error sources for the LCSRs. If using the $\phi_{2;\rho}^\|$ determined from the sum rules without the NLO-terms and the dimension-six condensates, we obtain $A_1(0)=0.230^{+0.028}_{-0.020}$, $A_2(0)=0.257^{+0.063}_{-0.043}$ and $V(0)=0.262^{+0.020}_{-0.016}$. Those values change from the above ones, i.e. Eqs.(\ref{F01},\ref{F02},\ref{F03}), determined from the sum rules with the NLO-terms and the dimension-six condensates by about $3\%-4\%$.

We put those TFFs versus $q^2$ in Fig.(\ref{fig_TFF}), where we have extrapolated them to all allowable $q^2$-region via the rapidly converging series in the parameter $z(t)$ expansion which is suggested by Refs.\cite{Khodjamirian:2010vf, Bourrely:2008za, Straub:2015ica}
\begin{eqnarray}
F_i(q^2) = \frac1{1-q^2/m_{R,i}^2}\sum_{k=0,1,2}a_k^i [z(q^2)-z(0)]^k,
\end{eqnarray}
where
\begin{eqnarray}
z(t)=\frac{\sqrt{t_+ - t}-\sqrt{t_+ - t_0}}{\sqrt{t_+ - t}+\sqrt{t_+ - t_0}}
\end{eqnarray}
with $t_\pm=(m_B\pm m_\rho)^2$ and $t_0=t_+(1-\sqrt{1-t_-/t_+})$. The values of the resonance masses $m_{R,i}$ can be found in Ref.~\cite{Straub:2015ica}, and $F_i$ stands for the three TFFs. The parameters $a_k^i$ are fixed such that $\Delta < 1\%$, which are put in Table~\ref{analytic}. The measure of the quality of extrapolation $\Delta$ is defined as
\begin{equation}
\Delta=\frac{\sum_t\left|F_i(t)-F_i^{\rm fit}(t)\right|} {\sum_t\left|F_i(t)\right|}\times 100, \label{delta}
\end{equation}
where $t\in[0,\frac{1}{2},\cdots,\frac{27}{2},14]{\rm GeV}^2$.

\begin{table}[htb]
\begin{center}
\begin{tabular}{c  c  | c }
\hline
\multicolumn{2}{c|}{Our prediction}  &  ~~~$3.19^{+0.65}_{-0.62}$~~~ \\ \hline
\multicolumn{2}{c|}{Omn\`{e}s parametrization~\cite{Flynn:2008zr}} & $2.80(20)$ \\ \hline
                                                    & LCSR~\cite{Ball:2004rg}   & $2.75(24)$ \\
\raisebox {1.5ex}[0pt]{BABAR~\cite{delAmoSanchez:2010af}}     & ISGW~\cite{Scora:1995ty}          & $2.83(24)$ \\ \hline
                                                    & LCSR~\cite{Ball:2004rg}  & $2.85(40)$ \\
\raisebox {1.5ex}[0pt]{BABAR~\cite{Aubert:2005cd}}     & ISGW~\cite{Scora:1995ty}         & $2.91(40)$ \\
\hline
\end{tabular}
\caption{The predicted $|V_{\rm ub}|$ in unit $10^{-3}$. The estimations of the Omn\`{e}s parametrization~\cite{Flynn:2008zr} and BABAR collaboration~\cite{delAmoSanchez:2010af,Aubert:2005cd} are also presented as a comparison.}\label{Gammatota}
\end{center}
\end{table}

We apply the extrapolated $B\to\rho$ TFFs for the semi-leptonic decays, $B^0\to\rho^-\ell^+\nu_\ell$ and $B^+\to\rho^0\ell^+\nu_\ell$. Their branching ratios and lifetimes are~\cite{Agashe:2014kda}: ${\cal B}(B^0\to\rho^-\ell^+\nu_\ell)=(2.94\pm0.21)\times 10^{-4}$ and $\tau(B^0)=1.520\pm0.004~{\rm ps}$; ${\cal B}(B^+\to\rho^0\ell^+\nu_\ell)=(1.58\pm0.11)\times 10^{-4}$ and $\tau(B^+)=1.638\pm0.004~{\rm ps}$. Those two semi-leptonic decays can be adopted to determine the CKM matrix element $|V_{\rm ub}|$, we present the results in Table~\ref{Gammatota}. Both of them lead to the same predictions, $|V_{\rm ub}| = 3.19^{+0.65}_{-0.62}$, where the errors are squared averages of the errors from $\xi_{(2,4);\rho}^\|$, the Borel window, the continuum threshold $s^{B}_0$ for the $B\to\rho$ TFFs, the $b$-quark mass, the $B$-meson decay constant and the uncertainties from the measured lifetimes and branching ratios, respectively. If using the $\phi_{2;\rho}^\|$ determined from the sum rules without the NLO-terms and the dimension-six condensates, we obtain $|V_{\rm ub}| = 3.36^{+0.66}_{-0.64}$, which changes from the one determined from the sum rules with the NLO-terms and the dimension-six condensates by about $5\%$. Table \ref{Gammatota} shows our result is consistent with the Omn\`{e}s parametrization and BABAR prediction within errors.

\section{Summary} \label{section:4}

The BFT provides a clean physical picture for the perturbative and non-perturbative properties of QCD and provides a systematic way to derive the SVZ sum rules for hadron phenomenology. In the paper, we have studied the moments of the $\rho$-meson leading-twist DA $\phi_{2;\rho}^\|$ via the SVZ sum rules up to dimension-six operators and by taking the NLO QCD correction to the perturbative part. Our predictions for the second and fourth moments $\langle \xi_{2;\rho}^\| \rangle$ and $\langle \xi_{4;\rho}^\| \rangle$, which lead to the Gegenbauer moments $a_{2;\rho}^{\|}|_{1{\rm GeV}} = 0.119(82)$ and $a_{4;\rho}^\| |_{1 {\rm GeV}} = -0.035(100)$. They indicate a double humped behavior for $\phi_{2;\rho}^\|$, which agrees with most of predictions done in the literature.

\begin{figure}[tb]
\centering
\includegraphics[width=0.45\textwidth]{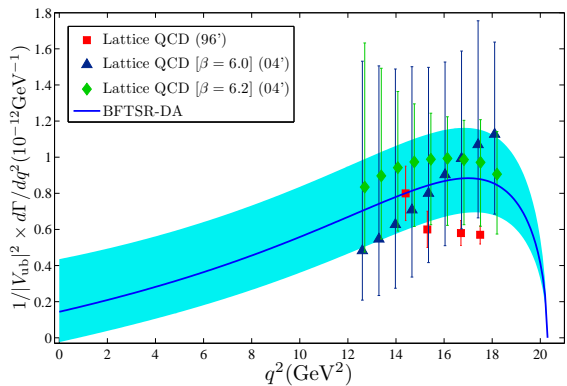}
\caption{The predicted differential decay width $1/|V_{\rm ub}|^{2} \times d\Gamma/dq^2$, where the shaded band shows the squared average of all the errors from the mentioned error sources. The lattice QCD predictions~\cite{Flynn:1995dc, Bowler:2004zb} are presented as a comparison. } \label{Fig:dGamma}
\end{figure}

The $\rho$-meson DA is a key component for $\rho$-meson involved high-energy processes. A better determination of the $\rho$-meson DA shall be helpful for a better understanding of the $\rho$-meson physics. As an application of $\phi_{2;\rho}^\|$, we calculate the $B\to\rho\ell\nu_\ell$ semi-leptonic decays within the LCSR via a chiral correlator. It is found that the extrapolated $B\to\rho$ TFFs agrees with the lattice QCD predictions~\cite{Flynn:1995dc, Bowler:2004zb} within errors. This can be more clearly shown by Fig.(\ref{Fig:dGamma}), which shows the different decay width $1/|V_{\rm ub}|^2\times d\Gamma/dq^2$. Our present obtained $\rho$-meson DA $\phi_{2;\rho}^\|$ shall be further constrained/tested by more data available in the near future, and we hope the definite behavior can be concluded finally.  \\

{\bf Acknowledgments}: This work was supported in part by the Natural Science Foundation of China under Grant No.11275280, No.11547305, No.11547015, and by Fundamental Research Funds for the Central Universities under Grant No.CDJZR305513.

\end{document}